# A novel scheme for simple and precise measurement of the complex refractive index and thickness of thin films

Yu Peng

*School of Physics, Beijing Institute of Technology, Beijing, 100081, P. R. China*

Email: pengyu@bit.edu.cn

We demonstrate applications of a novel scheme which is used for measuring refractive index and thickness of thin film by analyzing the relative phase difference and reflected ratio at reflection point of a monolithic folded Fabry-Perot cavity (MFC). The complex refractive index and the thickness are calculated according to the Fresnel formula. Results show that the proposed method has a big improvement in accuracy with simple and clear operating process compared with the conventional Ellipsometry.

## 1. INTRODUCTION

Ellipsometry is a versatile and powerful optical technique for the investigation of the complex refractive index or thickness, which gives access to fundamental physical parameters and is related to a variety of sample properties, including morphology, glass quality, chemical composition, and electrical conductivity.

The mechanism of Ellipsometry is based on exploiting the polarization transformation when a beam of polarized light is reflected from or transmitted through the interface or film and then obtains relative phase difference between S and P polarization components [1]. However, Ellipsometry [2-5] is applied with complicated measuring process and limited accuracy of thin film thickness accuracy, 1 nm, and refractive index accuracy, 0.01, due to big error when reading angle [6].

Here, we demonstrate a novel scheme which is used for measuring the thickness and refractive index of thin films by measuring the relative phase difference between S and P at oblique incidence point. Compared to Ellipsometry, this method can improve the accuracy of measuring film thickness up to 0.01 nm and refractive index accuracy, 0.0001, and be used with simple and clear operating process.

## 2. EXPERIMENTS AND DISCUSSION

In our experiments, a setup, shown in figure 1, is used to determine the thickness and refractive index of thin films by measuring the relative phase difference between S and P. In this setup, a commercial MQW GaAlAs laser diode (Hitachi HL6738MG) with spectral range from 680 nm to around 695 nm operate at 689 nm and provide an output power of 5.3 mW. The threshold current is 50 mA and the laser is driven at a pump current of 60 mA. The temperature of laser diode is stabilized at room temperature by thermoelectric cooler. The laser beam is incident on the diffraction grating of 2400 grooves/mm with spectral resolution of 50 GHz. Laser frequency of laser diode with a wide spectral range is selected with diffraction grating. And resonance signal appear when laser frequency is equal to resonant frequency of the MFC [7] by controlling PZT.

The confocal MFC made of optical quartz glass is schematically shown in figure 2, including two optical planes, S1, S2, and an optical spherical S3, which define a confocal F-P cavity. The thin film coated material is Tantalum oxide ($Ta_2O_5$), and the coupling plane S1 has a reflectivity of 0.91 (0.88) for S (P) polarization. Plane S2 is a total internal reflection surface, and S3 is a spherical mirror with a reflectivity of 0.999. The geometric length of the MFC is 30.17 mm, designed equal to the radius of curvature of S3. The finesse of the MFC is 33 (25) for S (P) polarization. The laser beam, with an external incidence angle of 45° at point A, travels along the route of ABCBADA. Resonance feedback beam is collinear with the incident light but in the opposite propagating direction, shown in figure 2.

In our experiments, the first medium is optically denser than the second, $n' = n_1 / n_3 = 0.687$ is relative index of refraction. $\theta_c = 86.83°$ is critical value, and $\theta_B = 69.00°$ is Brewster angle. In the case of the reflected wave, the phase changes of each component of the reflected wave will depend on the incidence magnitudes, $\theta$. To apply the Fresnel formula [8], the intensity of the light which is totally reflected is equal to the intensity of the incident light.

$$r_s = \frac{\cos\theta - i\sqrt{\sin^2\theta - n'^2}}{\cos\theta + i\sqrt{\sin^2\theta - n'^2}} = |r_s|e^{i\delta_s}$$

$$r_p = \frac{n'^2\cos\theta - i\sqrt{\sin^2\theta - n'^2}}{n'^2\cos\theta + i\sqrt{\sin^2\theta - n'^2}} = |r_p|e^{i\delta_p} \tag{1}$$

$$\tan(\delta_s / 2) = -\sqrt{\sin^2\theta - n'^2} / \cos\theta)$$

$$\tan(\delta_p / 2) = -\sqrt{\sin^2\theta - n'^2} /(n'^2 \cos\theta) \tag{2}$$

According equation (1) (2), we determine the changes $\delta_s$, $\delta_p$, in the phases of the components of the reflected and the incident wave. Write down an expression for the relative phase difference

$$\Delta_2 = \delta_s - \delta_p = 2*\tan^{-1}(\cos\theta\sqrt{\sin^2\theta - n'^2} / \sin^2\theta) \tag{3}$$

The measure setup is shown in figure 1, for obtaining relative phase difference $\Delta$ inside optically denser medium. S and P resonance are separately acquired by adjusting the half-wave plate, which is used for changing the polarization direction of incidence to S and P respectively .

Figure 3 shows the changing procedure of S (green) and P (blue) component resonance of confocal MFC. In this figure, 3.41(3) GHz of the free spectral range is obtained by scanning laser frequency, in which higher order mode appear in the middle of the spectral range because of incidence pattern dismatching to the confocal MFC. S and P resonance are separately acquired by adjusting the half-wave plate in figure 1, which is used for changing the polarization direction of incidence. Total relative phase difference of round trip is determined by [9]

$$\Delta = 2*\pi*\Delta v*t \tag{4}$$

Where $t$ is propagation time, determined by $t = n*2*l/c$, in which n, $l$, c represent index of refraction of MFC material, geometric length of MFC cavity and speed of light in vacuum respectively. According to figure 3, adjacent axial modes of S and P resonances of MFC, $\Delta v_0$ is 1.03(0) GHz spectral shift, therefore frequency difference between S and P polarization is determined by $\Delta v = N*FSR \pm \Delta v_0$, in which N represents the number that how many FSRs the $\Delta v$ has, and FSR represents free spectrum range with the value of 3.41(3) GHz for the case of confocal MFC. We roughly estimate $\Delta v = N*FSR + \Delta v_0$ and $\Delta v = N*FSR - \Delta v_0$ respectively, and determine which expression is proper to ensure that expression (4) and (5) are same quantity by integer N and theoretical value of the relative phase difference at reflection point A. By equation (4), we can get the value of total relative phase difference, $\Delta$. On the other hand, total relative phase difference is expressed by [8]

$$\Delta = 2*\Delta_1 + 2*\Delta_2 + \Delta_3 + \Delta_4 \tag{5}$$

Where $\Delta_1$, $\Delta_2, \Delta_3$, and $\Delta_4$ are relative phase difference between S and P resonances caused by point A, B, C, and D respectively. For the total internal reflection point B, $\Delta_2$ with the value of 43.49° is calculated by the equation (3). At normal incidence point C,D, around $\Delta_3 \approx \Delta_4 \approx -180.00°$ of relative phase difference are obtained. According to equation (5), therefore the relative phase difference at reflection point A, $\Delta_1$ is inferred to be around -167.42°, where contains some deviation with -180.00°. The deviation, we think, is caused by coating films of plane S1, shown in figure 4.

In thin film area, the relative phase difference, $\Delta_1$, and amplitude ratio changes $\tan\psi$ are the key elements to measure refractive index $n_2$ and thickness $d_2$ of a film. Ellipsometry [2-5] is applied not only with complicated measuring process, but also with limited accuracy of thin film thickness, 1 nm and refractive index accuracy, 0.01, due to big error when reading angle [6]. Therefore, we propose this method, which can improve the measuring accuracy of film thickness up to 0.01 nm and accuracy of refractive index 0.0001, and be used with simple and clear operating process.

We magnify the S1 of the confocal MFC in Fig.5, which depicts a typical system consisting of a film of refractive index $n_2$ and thickness $d_2$ on a reflecting substrate of index $n_3$ immersed in a medium of index $n_1$. Reflection causes a change in the relative phases of the S and P waves and a change in the ratio of their amplitudes.

The ratio of the parallel and normal reflection coefficients is defined as $\rho$, where this ratio may be expressed in terms of $\Delta_1$ and $\psi$ as

$$\rho = R^s/R^p = \tan\psi \exp(j\Delta_1) \tag{6}$$

Reflected light is characterized by $\Delta_1$, defined as the change in phase, and the angle $\psi$, the arctangent of the factor by which the amplitude ratio changes. The complex refractive index (RI) can be defined in terms of a real part and an imaginary part as $n = n_r + ik$, where k means the extinction coefficient.

The total reflection coefficients $R^s$ and $R^p$, which include the contributions of reflections from lower boundaries, are given by references[8-9]

$$\begin{cases} R^s = \dfrac{r_{12}^s + r_{23}^s \exp D}{1 + r_{12}^s r_{23}^s \exp D} \\ R^P = \dfrac{r_{12}^p + r_{23}^p \exp D}{1 + r_{12}^p r_{23}^p \exp D} \end{cases} \quad (7)$$

The normal (s) and parallel (p) reflection coefficients for light incident at the immersion film interface, $r_{12}^p, r_{23}^p, r_{12}^s, r_{23}^s$ are

$$\begin{cases} r_{12}^p = (n_2 \cos\phi_1 - n_1 \cos\phi_2)/(n_2 \cos\phi_1 + n_1 \cos\phi_2) \\ r_{23}^p = (n_3 \cos\phi_2 - n_2 \cos\phi_3)/(n_3 \cos\phi_2 + n_2 \cos\phi_3) \\ r_{12}^s = (n_1 \cos\phi_1 - n_2 \cos\phi_2)/(n_1 \cos\phi_1 + n_2 \cos\phi_2) \\ r_{23}^s = (n_2 \cos\phi_2 - n_3 \cos\phi_3)/(n_2 \cos\phi_2 + n_3 \cos\phi_3) \\ n_1 \sin\phi_1 = n_2 \sin\phi_2 = n_3 \sin\phi_3 \end{cases} \quad (8)$$

and

$$D = -4\pi j d_2 n_2 \cos\phi_2 / \lambda \quad (9)$$

Where $\lambda$ is the wavelength of light used, $j = -1^{1/2}$ and $d_2$ is the film thickness. Considering light incident (at angle $\phi_1$) at the boundary between the immersion medium and film, the cosine of the angle of refraction can be written

$$\cos\phi_2 = \left\{1 - \left[(n_1/n_2)\sin\phi_1\right]^2\right\}^{1/2} \quad (10)$$

According to Eqn (6-10), as a result, the complex refractive index of the reflecting surface is expressed

$$n_2 = N - iNK \quad (11)$$

Where N and K can be written

$$\begin{cases} N = n_1 \sin\phi_1 tg\phi_1 \cos 2\Psi / (1 + \sin 2\Psi \cos \Delta_1) \\ K = tg 2\Psi \sin \Delta_1 \end{cases} \quad (12)$$

The $\Delta_1$ can be gotten by our setup above and $\Psi$ can be calculated by Eqn(6) with the ratio of the parallel and normal reflection coefficients. The value of the amplitude ratio changes, $\tan\psi$, can be calculated to be 0.91(8) by S and P resonances of MFC according Fig.3. The uncertainty of $\tan\psi$ is determinged by power amplitudes fluctuations of the P and S polarized signals, which is associated with fluctuations in the laser power output. Therefore, according to Eqn (11-12), complex refractive index $n_2$ can be calculated to be 2.183(1)-1.326(4) i. And finally, film thickness $d_2$ values can be calculated to be 162.0(1) nm according to Eqn (9).

## 3. CONCLUSION

In summary, we design a novel setup to precisely determine the relative phase difference $\Delta_1$, and by roughly given the amplitude ratio changes $\tan\psi$, complex refractive index $n_2$ and film thickness $d_2$ can be calculated to be 2.183(1)-1.326(4) i, and 162.0(1) nm according the theory above separately. The above analyses have shown that the scheme used to measure the complex refractive index of films is highly accurate compared to conventional Ellipsometry. It also can be used to measure the refractive index of other materials. In the future we envisage a polychromatic version of the setup using a tunable laser or multiple laser sources of different wavelengths to validate the method.

## 4. ACKNOWLEDGMENTS

The author thanks Dr. Erjun Zang (National Institute of Metrology, China) for his useful discussions, especially his designs for MFC. This work is supported by Basic Research Funds from Beijing Institute of Technology (Grant No. 20121842004).

# REFERENCES

[1] R. M. A. Azzam and N. M. Bashara, Ellipsometry and Polarized Light (1987).

[2] K. Riedling, Ellipsometry for Industrial Applications (New York, Springer-Verlag) (1988).

[3] R. M. A. Azzam, Optics Letters **10**(7) 309-311(1985).

[4] R. M. A. Azzam, T. L. Bundy, and N. M. Bashara, Opt. Commun. **7** 110-115(1973).

[5] R. M. A. Azzam and N. M. Bashara, Opt. Commun. **5** 5-9(1972).

[6] Hiroyuki Fujiwara, Spectroscopic Ellipsometry: Principles and Applications, (John Wiley and Sons) (2007).

[7] Yang Zhao, Yu Peng, Tao Yang, Ye Li, Qiang Wang, Fei Meng, Jianping Cao, Zhanjun Fang, Tianchu Li, and Erjun Zang, Optics Letters, 36(1), pp. 34-36 (2011)

[8] Max Born and Emil Wolf, Principles of Optics 43-57 (2003).

[9] F. A. Jenkins and H. E. White, Fundamental of physical optics (1937).

**Figure captions:**

**Figure 1.** (Color online) The setup for obtaining relative phase difference: LD, laser diode; HWP, half-wave plate; PZT, piezoelectric transducer; MFC, monolithic folded F-P confocal cavity; PD, photodetector; PBS, polarizing beam splitter; blue arrow, direction of incidence's polarization; dotted red arrow, resonant feedback

**Figure 2**. (Color online) Dimensional map and Photograph of the confocal MFC.(left); Scheme of resonant feedback. Perpendicular(S) polarization points out of the plane of incidence (green point). Parallel (P) polarization lies parallel to the plane of incidence (blue stick).(right)

**Figure 3.** (Color online) Line shape changing procedure of S (green) and P (blue) component resonance of confocal MFC. Adjacent axial modes of S and P resonances of MFC, $\Delta v_0$ is 1.030 GHz spectral shift

**Figure 4.** (Color online) The relative phase difference between S and P versus incidence angles. Red mark point, A,B,C,D represent corresponding resonant reflection point inside the MFC respectively. The relative phase difference at reflection point A is inferred to be around -167.4°

**Figure 5.** (Color online) Reflection and transmission from the film-covered surface. Perpendicular(S) polarization points out of the plane of incidence (green point). Parallel (P) polarization lies parallel to the plane of incidence (blue stick).

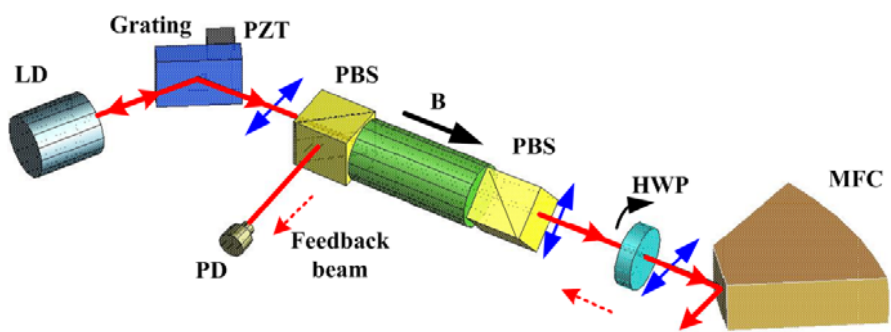

**Figure 1.**

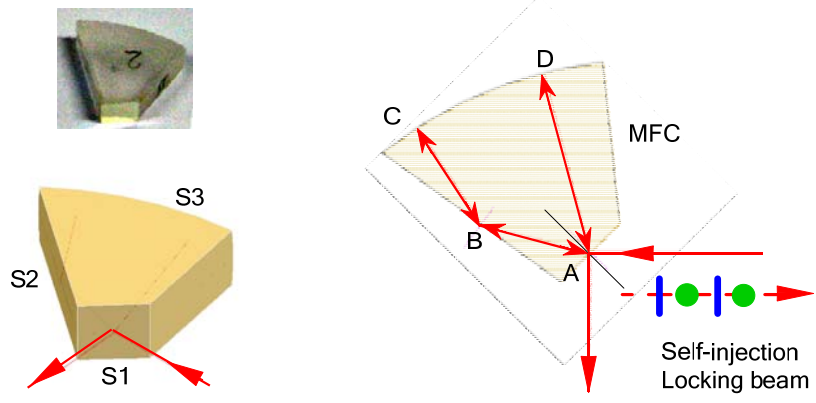

**Figure 2.**

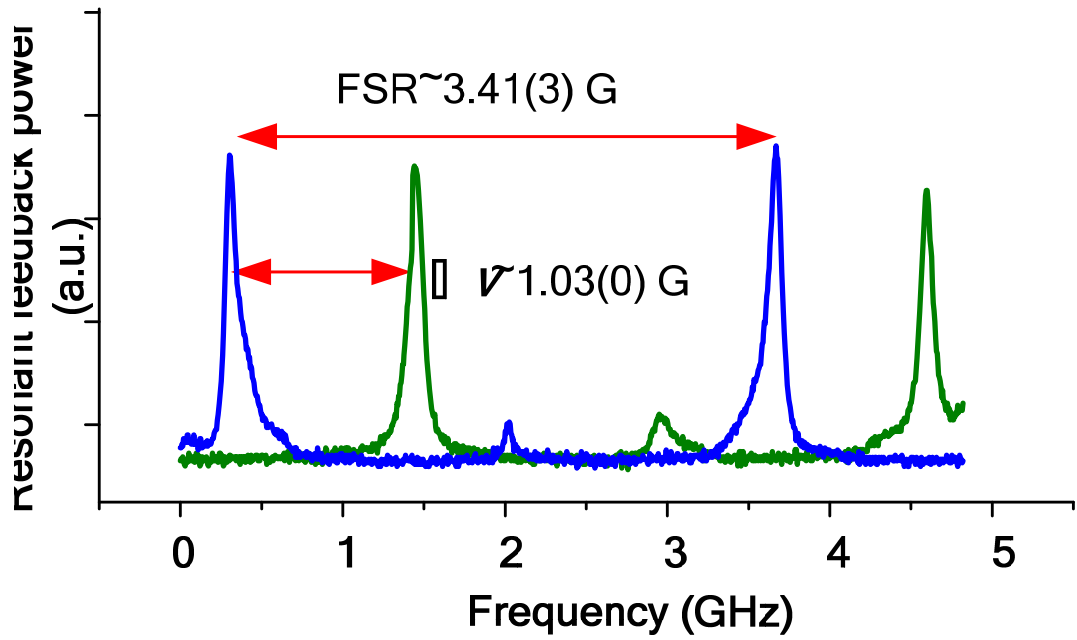

Figure 3.

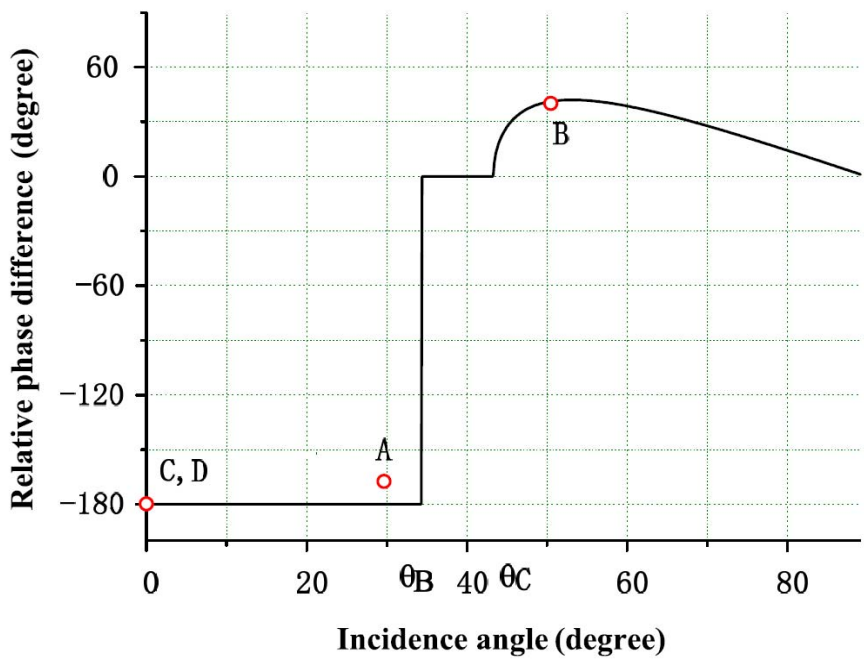

Figure 4.

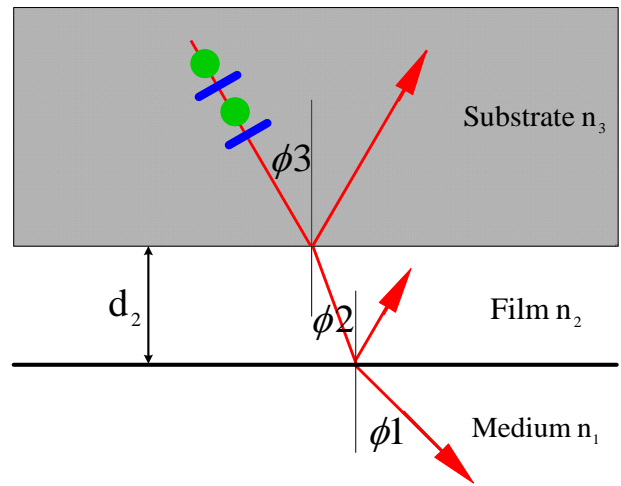

**Figure 5.**